\documentclass{JHEP3}

\usepackage{amsmath,amssymb,epsfig}
\usepackage{graphicx}
\newcommand{\amp}{&}
\newcommand{\volu}{\rho}
\newcommand{\voluume}{\rho}

\newcommand{\dil}{\tau}
\newcommand{\dilaton}{\tau}

\newcommand{\md}{\phi_i}
\newcommand{\mpl}{\bar{m}_{\mbox{\tiny{P}}}}

\preprint{MIT-CTP-3905, SLAC-PUB-12999\\ {\tt 0711.2512 [hep-th]}}

\title{Inflationary Constraints on Type IIA String Theory}

\author{Mark P.  Hertzberg$^1$\footnote{Electronic address: {\tt mphertz} at {\tt mit.edu}},
Shamit Kachru$^2$\footnote{Electronic address: {\tt skachru} at {\tt stanford.edu}}, 
Washington Taylor$^1$\footnote{Electronic address: {\tt wati} at {\tt mit.edu}}, 
and Max Tegmark$^1$\footnote{Electronic address: {\tt tegmark} at {\tt mit.edu}} 
\\
~\\
$^1$Dept.~of Physics, Massachusetts Institute of Technology, Cambridge, MA 02139, USA\\
$^2$Dept.~of Physics and SLAC, Stanford University, Stanford, CA 94305, USA \\~\\
}

\abstract{

We prove that inflation is
forbidden in the most well understood class of semi-realistic type IIA string
compactifications: Calabi-Yau compactifications with only standard NS-NS
3-form flux, R-R fluxes, D6-branes and O6-planes at large volume and small string coupling.
With these ingredients, the first slow-roll parameter satisfies
$\epsilon\ge\frac{27}{13}$ whenever $V>0$, ruling out both
inflation (including brane/anti-brane inflation) and de Sitter vacua in this limit.  
Our proof is based on the dependence of the
4-dimensional potential on the volume and dilaton moduli in the
presence of fluxes and branes. 
We also describe broader classes of IIA models which 
may include cosmologies with inflation and/or de Sitter vacua.
The inclusion of extra ingredients, such as NS 5-branes
and geometric or non-geometric NS-NS fluxes, evades the
assumptions used in deriving the no-go theorem. We focus on NS
5-branes and outline how such ingredients may prove fruitful for
cosmology, but we do not provide an explicit model.  We contrast the
results of our IIA analysis with the rather different situation in IIB.

}

\maketitle

\begin{document}

\newpage

\section{Introduction}

Our desire to understand the large-scale properties of our Universe is
one of the motivations for studying fundamental microphysical
theories such as string theory.  
Indeed, observing the early Universe may be our most promising path toward confronting 
string theory with data.
The leading paradigm for explaining the large-scale isotropy, homogeneity,
and flatness of the Universe, as well as its $\mathcal{O}(10^{-5})$ seed
fluctuations, is cosmological inflation 
\cite{
Guth,Linde,AlbrechtSteinhardt82,Linde83}.
Specifically, by assuming that there exist one or more scalar fields that undergo
slow rolling in the early Universe in a potential energy function of
just the right shape, one can explain these large-scale properties and
predict the numerical values of as many as eight cosmological parameters, many of which 
have now been accurately measured \cite{8pars,WMAP06,SDSS06}.
The leading candidate for a fundamental
microphysical theory is string theory, and so we would like to know
how generically string theory can accommodate such potential energy
functions.

It is rather well known that the conditions for inflation do not arise
easily in string theory, or in other words, that a {\em generic} point
in field space may not be expected to satisfy the slow-roll conditions
\cite{Hellerman, Fischler, Brustein}.  In part this is because of
issues like the $\eta$ problem (essentially, that the potential varies too quickly),
which also complicate attempts to build
inflationary models in quantum field theory and supergravity theories.
There are, however, three reasons to suspect a priori that string
theory can accommodate inflation: Firstly, the potential energy is
typically a function of hundreds of fields, which means that there is
a  large field 
space to explore.  Secondly, there are an
exponentially large number of infinite families of potential energy functions, parameterized
by typically hundreds of discrete fluxes in the compact space \cite{fluxrevs1,fluxrevs2,fluxrevs3,fluxrevs4,fluxrevs5}.
Thirdly, there are at least many millions of different topologies, such as Calabi-Yau manifolds,
that in general give rise to qualitatively different
physical theories in 4 dimensions.  It is reasonable to suspect that
occasionally in this vast space of possibilities, the conditions for
inflation are satisfied.

For this reason, the past few years have seen intense investigation into the
possibility of inflation driven by closed string moduli \cite{modulione,modulitwo,
modulithree,modulifour,modulifive}, axions 
\cite{axionone,axiontwo,axionthree,axionfour,axionfive}, or brane 
positions \cite{braneone,braneshiu,branetwo,branethree,branefour,branefive,branesix,braneseven} in 
the extra dimensions.  
In the most intensely studied
case of IIB string compactifications on Calabi-Yau orientifolds,
the conclusion at this point is that one can probably build working
models, at the cost of fine-tuning the relevant potentials.  
Some of these models could even have interesting observable signatures \cite{strings,NG}.
Reviews of this general subject appear in \cite{reviews1,reviews2,reviews3,reviews4,reviews5}.

One feature of the existing constructions is that they are implicit, relying
at some point on either non-compact models of regions of the compactification
space, or on the ability to perform tunes which (though seemingly possible based
on detailed theoretical considerations) are not performed explicitly.
One would ideally like to build simpler models, where all of the calculations
are performed in a completely explicit and reliable way.
Commonly, compactifications suffer from
unstabilized moduli in the low energy description, or the
calculations stabilizing moduli  apply in a regime that, while apparently
numerically controlled,
is not under parametric control.\footnote{This in particular applies in any 
class of stabilized compactifications 
where the number of choices of fluxes, branes, etc., while perhaps very large, is ${\it finite}$.  
In such models, there is perforce a limit on how small $g_s$ can be, namely the smallest $g_s$
obtained in the finite list.  Of course if the finite number is sufficiently large, the smallest attainable coupling may be quite small, so this may
not be a serious limitation.}

In some limits of string theory, however,  we now have explicit
examples of stabilized models with parametric control of the moduli
potential.  The best understood case occurs in massive IIA string
theory; namely IIA string theory with R-R 0-form flux, compactified on
a Calabi-Yau orientifold.  The 10-dimensional massive IIA supergravity
action was suggested in \cite{Romans}.  The compactification of this
theory on a Calabi-Yau orientifold was performed in a 4-dimensional
supergravity formalism in \cite{Grimm} and the stabilization was
obtained in \cite{DGKT}.\footnote{With additional ingredients (geometric flux)
 stabilization was achieved in \cite{vz}.}  
 The 10-dimensional description of these compactifications was further studied in \cite{Acharya}.
 Since these models carry at most ${\cal N}=1$ supersymmetry in 4 dimensions,
 and gauge groups and chiral matter can be incorporated in this context, we consider them
 to be at least semi-realistic.

An investigation of cosmology in these IIA compactifications was
initiated in Ref.~\cite{Hertzberg} by considering some specific simple
examples and showing that inflation could not occur in these examples.
In the present work we extend that study.  We use a simple scaling
analysis of various terms which appear in the low-energy 4-dimensional
potential to rule out inflation and de Sitter vacua at large volume
and weak coupling in all IIA Calabi-Yau models with conventional fluxes,
D-branes, and O-planes. 
This means that inflation imposes the constraint that our Universe
is not in this portion of the landscape.
 We emphasize that the derivation of this no-go result is only
valid in the large volume limit and can be evaded by various other
structures including NS 5-branes as well as geometric and
non-geometric NS-NS fluxes, indicating that IIA compactifications
containing these ingredients may be a good place to look for string
models with inflation and/or de Sitter vacua.  Indeed, as we were
completing this work we received a copy of \cite{Eva}, in which more
explicit IIA de Sitter models are constructed by using geometric NS-NS
fluxes, 5-branes, and various other ingredients.

The structure of this article is as follows:  In Section \ref{SR} we
summarize the IIA supergravity theory in 10 dimensions and outline the
dimensional reduction to 4 dimensions.  We explain the key step in the
analysis of the paper, which involves considering 2-dimensional slices
in the full moduli space parameterized by the volume and dilaton
moduli of the compactification.  The behavior of the four (space-time)
dimensional potential on these 2-dimensional slices of moduli space
allows us to place a lower bound on the slow-roll parameter
$\epsilon$.  In Section \ref{Nogo} we compute the scaling of the
various terms appearing in the 4-dimensional potential energy function
$V$ in terms of the two model-independent moduli. 
We prove that both inflation and de Sitter vacua are forbidden at
large volume and weak string coupling
when standard fluxes,
D6-branes, and O6-planes are included; the slow-roll parameter is
bounded below in this case by $\epsilon\ge\frac{27}{13}$ whenever $V>0$.  In Section
\ref{Evade} we describe some additional ingredients such as NS
5-branes, geometric fluxes and non-geometric NS-NS fluxes which can be
included in type IIA and which lead to terms in the 4-dimensional potential
with scaling properties allowing us to evade the no-go theorem.  In
Section \ref{sec:IIB} we discuss the type IIB theory.  We show how the
structure of the IIB theory differs from the IIA theory from the point
of view taken in this paper and discuss the connection of our IIA
results with previous work on inflation in IIB models.  We
discuss our results in Section \ref{Discussion}.
More details regarding the kinetic energy and potential energy are provided in Appendix \ref{App}.

\section{Type IIA Compactifications}\label{SR}

We investigate large volume and small string coupling compactifications where it
is valid to perform computations using supergravity.
We study the 10-dimensional type IIA supergravity theory,
where we include conventional NS-NS and R-R field strengths, 
as well as  D6-branes and O6-planes:
\begin{eqnarray}
S \amp=\amp\frac{1}{2 \kappa_{10}^2}\int d^{10}\!x\sqrt{-g}\,
e^{-2\phi}
\Big{(}R+4(\partial_{\mu}\phi)^2
-\frac{1}{2} | H_3 |^2 - e^{2\phi}\sum_{p}|F_p|^2\Big{)}\nonumber\\
\amp\amp -\mu_6\int_{\mbox{\tiny{D6}}}d^7\xi\sqrt{-g}\,e^{-\phi}
+ 2\mu_6\int_{\mbox{\tiny{O6}}}d^7\xi\sqrt{-g}\,e^{-\phi}
\label{gravity1} \end{eqnarray}
where $R$ is the 10-dimensional Ricci scalar, $\phi$ is the scalar
dilaton field, $H_3$ is the NS-NS 3-form field strength that is
sourced by strings, $F_p$ are the R-R $p$-form field strengths
($p=0,2,4,6$) that are sourced by branes, $\kappa_{10}^2 = 8\pi
G_{10}$ is the gravitational strength in 10-dimensions, and $\mu_6$
($-2\mu_6$) is the D6-brane (O6-plane) charge and tension.  
We have set all fermions to zero as we are interested in solutions with maximal space-time symmetry.
There are also Chern-Simons contributions to the action.  These are essentially
topological, and are independent of the dilaton as well as the overall
scale of the metric (in string frame).  We expect that as in
\cite{DGKT} the contribution to the action from the Chern-Simons terms
will vanish on-shell,\footnote{More precisely, integrating out the 4-dimensional
non-dynamical field $dC_3$ as a Lagrange multiplier gives an equation
which must be satisfied by the axions, but the Chern-Simons terms do not
otherwise affect the 4-dimensional potential.} so that we need not consider it
further here.  Although there are some subtle questions regarding the
definition of orientifolds in these massive IIA backgrounds and
whether these backgrounds can be described in a weak coupling string
expansion \cite{Moore,Banks}, these compactifications seem to be
described adequately in the 4-dimensional supergravity formalism of \cite{Grimm},
corresponding to the 10-dimensional massive supergravity analysis when the
sources are uniformly distributed in the compactification space.

\subsection{Compactification}

We now perform a Kaluza-Klein compactification of this theory from
10 dimensions to 4 dimensions.  Let us first focus on the gravity sector.
Assuming that we can neglect any dependence of the Ricci scalar on the
compact space coordinates, we can
integrate over the compact space, giving
\begin{equation}
\int d^{10}\!x\sqrt{-g_{10}}\,e^{-2\phi}R = 
\int d^{4}\!x\sqrt{-g_{4}}\,\mbox{Vol}\,e^{-2\phi}R
\end{equation}
where Vol is the 6-dimensional volume of the compact space.  The
volume, dilaton, and all the other fields that describe the size and shape
of the compact space are scalar fields in the 4-dimensional
description, known as moduli.   In addition to kinetic energy terms, the remaining terms in the
supergravity action, when reduced to 4 dimensions, describe a
potential function $V$ which depends on the moduli and fluxes in
any given model.

The key observation of this paper is that by studying the dependence
of the potential energy function $V$ on only two of the moduli, we
can learn a great deal about the structure of the potential relevant
for the possibility of inflation.  We define
the volume modulus of the compact space $\volu$ and
the dilaton modulus $\dil$ by\footnote{Note that we have
defined the volume modulus in the string frame, since this definition
relates to the K\"ahler moduli in the IIA theory.  
This differs from the conventional definition of 
K\"ahler moduli in analogous IIB orientifolds, where the Einstein
frame metric is used in defining the chiral multiplets.}
\begin{equation}
\volu\equiv(\mbox{Vol})^{\frac{1}{3}},\,\,\,\,\,\,
\dil\equiv e^{-\phi}\sqrt{\mbox{Vol}}.
\end{equation}
While $V$ depends on all moduli, we can explore the behavior of this
function on the whole moduli space by considering 2-dimensional slices of the
moduli space where all moduli other than $\dilaton$ and $\voluume$ are
fixed.  By showing that $V$ has a large gradient in the
$\dilaton$-$\voluume$ plane on every slice wherever $V$ is positive,
we will be able to rule out inflation on the {\em entire} moduli space, regardless of which fields we would like to identify as the inflaton.

In order to bring the gravity sector into canonical form, we perform a conformal transformation on the metric to the so-called Einstein frame:
\begin{equation}
g_{\mu\nu}^{E}\equiv \frac{\dil^2}{\mpl^2\,\kappa_{10}^2} g_{\mu\nu}^4,
\label{Einconf}\end{equation}
where $\mpl=1/\sqrt{8\pi G}\approx2\times 10^{18}$\,GeV is the
(reduced) Planck mass and $G$ is the 4-dimensional Newton constant.  
By re-expressing $R$ in terms of a 4-dimensional Ricci scalar and performing the conformal transformation, one finds that the gravity sector is canonical and that the fields $\volu$ and $\dil$ carry kinetic energy that is diagonal.  Although $\volu$ and $\dil$ do not have canonical kinetic energies, they are related to fields which do:
\begin{equation}
\hat{\volu}\equiv \sqrt{\frac{3}{2}}\,\mpl \ln\volu,\,\,\,\,\,\,
\hat{\dil}\equiv \sqrt{2}\,\mpl \ln\dil.
\end{equation}
Altogether we obtain the effective Lagrangian in 4 dimensions in the Einstein frame
\begin{equation}
\mathcal{L} = \frac{1}{16\pi G}R_E - \left[\frac{1}{2}(\partial_\mu\hat{\volu})^2 +\frac{1}{2}(\partial_\mu\hat{\dil})^2
 +\ldots\right] - V
\label{KE}\end{equation}
where the Ricci scalar $R_E$ and all derivatives are defined with
respect to the Einstein metric.  The dots indicate further kinetic
energy terms from all the other fields of the theory associated with
the compactification ($\phi_i$): the so-called K\"ahler moduli, complex structure
moduli, and axions\footnote{The axions arise from zero-modes of the
various gauge fields.}.  
The important point is that their contributions will always be {\em positive}.  
More details are provided in Appendix \ref{AppK}.
All contributions from the field strengths and D-branes \&
O-planes from eq.~(\ref{gravity1}) are described by some potential
energy function $V$.

\subsection{The Slow-Roll Condition}

From this action we can derive the slow-roll conditions on the
potential $V$ for inflation.  In order to write down the conditions in
detail we would need to know the precise form of the kinetic energy
with respect to all the moduli.  This can be done cleanly in the
4-dimensional supergravity formalism, as mentioned in Appendix \ref{AppK}.  
The important point here is that the first slow-roll parameter
$\epsilon$ involves partial derivatives of the potential with respect
to each direction in field space, and that 
the contribution from $\hat{\volu},\,\hat{\tau},\,\phi_i$ is non-negative.
In fact it is roughly the square of
the gradient of $\ln V$.  The contributions from $\hat{\volu}$ and
$\hat{\dil}$ thus give the following lower bound:
\begin{equation}
\epsilon \ge \frac{\mpl^2}{2}\left[\left(\frac{\partial \ln V}{\partial\hat{\volu}}\right)^2+
\left(\frac{\partial \ln V}{\partial\hat{\dil}}\right)^2\right].
\label{epsilon}\end{equation}
A {\em necessary} condition for inflation is $\epsilon\ll 1$ with $V>0$.  We will now prove that this condition is 
impossible to satisfy in this IIA framework, at large volume and weak coupling where our calculations apply.

\section{No-Go Theorem}\label{Nogo}

Having discussed the gravity and kinetic energy sector of the theory,
let us now describe the form of the potential energy function $V$.  By
using the bound on $\epsilon$ we will then prove that inflation is
forbidden for any Calabi-Yau compactification of type IIA string
theory in the large volume and small string coupling limit, when only
conventional NS-NS and R-R field strengths, D6-branes and O6-planes
are included.

\subsection{Potential Energy}

The potential energy arises from the dimensional reduction of the
terms in (\ref{gravity1}) associated with the various field strengths
$H_3$ \& $F_p$ ($p=0,2,4,6$) and the D6-branes \& O6-planes.  Let us
focus on some field strength $F_p$.  Such a field can have a
nonvanishing integral over any closed $p$-dimensional internal
manifold (homology cycle) of the compact space, and must satisfy a
generalized Dirac charge quantization condition:
\begin{equation}
\int_\Sigma F_p \propto f_\Sigma,
\end{equation}
where $f_\Sigma$ is an integer associated with number of flux quanta
of $F_p$ through each $p$-dimensional cycle $\Sigma$.  By choosing
different values for $f_\Sigma$ over a basis set of $p$-cycles, one
obtains a {\em landscape} of possible allowed potential energy
functions $V$.

The energy arising from a $p$-form flux $F_p$ comes from a term in
(\ref{gravity1}) proportional to $| F_p |^2$; since the $p$-form
transforms as a covariant $p$-tensor ({\it i.e.}, has  $p$ lower
indices), we contract with $p$ factors of $g^{\mu \nu}_6$, so that
\begin{equation}
|F_p|^2\propto \volu^{-p}
\end{equation}
in the string frame.  By including the appropriate factors of the volume and dilaton from the compactification and performing the conformal transformation to the Einstein frame, we have
the following contributions to $V$:
\begin{eqnarray}
V_3 \amp\propto\amp \volu^{-3} \dil^{-2} \,\,\,\,\,\,\,\,\,\mbox{for $H_3$}, \nonumber\\
V_p \amp\propto\amp \volu^{3-p} \dil^{-4}\,\,\,\,\,\,\,\mbox{for $F_p$}.
\end{eqnarray}
We also need the contribution from D6-branes and O6-planes.  In the
Einstein frame they scale as
\begin{eqnarray}
V_{\mbox{\tiny{D6}}}\amp\propto\amp \dil^{-3}\,\,\,\,\,\,\,\,\,\,\,\mbox{for D6-branes},\nonumber\\
V_{\mbox{\tiny{O6}}}\amp\propto\amp -\dil^{-3}\,\,\,\,\,\,\,\mbox{for O6-planes},
\end{eqnarray}
where we have indicated that O6-planes provide a negative
contribution, while all others are positive.

Altogether, we have the following expression for the scalar potential in 4-dimensions
\begin{eqnarray}
V \amp = \amp V_3+\sum_p V_p+V_{\mbox{\tiny{D6}}}+V_{\mbox{\tiny{O6}}} \nonumber \\
\amp = \amp \frac{A_3(\md)}{\volu^3\dil^2}+\sum_p\frac{A_p(\md)}{\volu^{p-3}\dil^4}
+\frac{A_{\mbox{\tiny{D6}}}(\md)}{\dil^3}-\frac{A_{\mbox{\tiny{O6}}}(\md)}{\dil^3}.\,\,\,\,\,\,\,
\label{potential}\end{eqnarray}
Here we have written the various coefficients as $A_j$ ($\ge0$), which
in general are complicated functions
of all the other fields of the theory $\md$, namely the remaining set
of K\"ahler moduli, complex structure moduli, and axions.  The
coefficients $A_j$ also depend on the choice of flux integers $f_\Sigma$.
This means that $V$ is in general a function of hundreds of fields,
for each of the exponentially large number of infinite families of possible flux
combinations on each of the many available Calabi-Yau manifolds.  
We have simply described its dependence on two of the fields: $\volu$
and $\dil$.  
An alternative proof of eq.~(\ref{potential}) from the perspective of
the 4-dimensional supergravity formalism is given in Appendix \ref{AppW}.
As discussed in Ref.~\cite{DGKT} this potential ensures that there exist
special points in field space which stabilize all the geometric moduli and
many axions.
We are interested, however, in exploring the full moduli space in search of inflation.

\subsection{Proof of No-Go Theorem}

We find that for any potential of the class we have
constructed so far, inflation is impossible anywhere in the moduli
space. The proof of this result is quite simple.  The key is to
observe that the potential in eq.~(\ref{potential}) satisfies
\begin{eqnarray}
-\volu \frac{\partial V}{\partial\volu}-3\dil\frac{\partial V}{\partial\dil}\amp=\amp 9 V+\sum_p p\,V_p 
\nonumber\\
\amp \ge \amp 9 V,
\label{operator}\end{eqnarray}
where the inequality comes from the fact that all $V_p\ge0$
($A_p\ge0$) and $p\ge0$.    Now, assuming we are in a region where
$V>0$, which is necessary for inflation, we can divide both sides by
$V$ and rewrite this inequality in terms of $\hat{\volu}$ and
$\hat{\dil}$ as
\begin{eqnarray}
\mpl\left| \sqrt{\frac{3}{2}}\left(\frac{\partial \ln V}{\partial\hat{\volu}}\right)+ 
3\sqrt{2}\left(\frac{\partial \ln V}{\partial\hat{\dil}}\right) \right| \ge 9.
\label{bound}\end{eqnarray}
This implies that it is impossible for both terms in eq.~(\ref{epsilon}) to be simultaneously small.
Specifically, comparing this inequality with eq.~(\ref{epsilon}), we see that $\sqrt{2\epsilon}/\mpl$ is
the distance to the origin on the plane spanned by 
$(\partial \ln V/\partial\hat{\volu},\partial \ln V/\partial\hat{\dil})$, 
while a sloped band around the origin is forbidden, implying the existence of a lower bound on $\epsilon$.
By minimizing $\epsilon$ subject to the constraint (\ref{bound}), we find the bound on the slow-roll parameter $\epsilon$ to be 
\begin{equation}
\epsilon\ge\frac{27}{13}\,\,\,\,\mbox{whenever}\,\,V>0.
\label{QED}\end{equation}
Hence both inflation and de Sitter vacua are forbidden everywhere in
field space.  

Indeed in any vacuum $\partial V/\partial \rho =
\partial V/\partial \tau = 0$, so eq.~(\ref{operator}) implies $V = -(\sum p V_p)/9$.
By assuming $V_p>0$ for at least one of $p=2,\,4$, or 6, then Minkowski vacua are forbidden 
also.\footnote{We almost certainly need $V_p>0$ for at least one of $p=2,\,4$, or 6 in order to be in the large volume and small string coupling limit, since the 3-form, 0-form, and number of D6/O6 planes are tightly constrained by a tadpole condition.}
This type of relation was used in \cite{DGKT} to show that vacua in a
specific IIA compactification must be anti-de Sitter, and in \cite{Kallosh} to
rule out a simple F-term uplift of this model; here we have shown that
this relation holds very generally in IIA compactifications, and
furthermore rules out inflation anywhere on moduli space for
potentials containing only terms of the form (\ref{potential}).
For stabilized compactifications, the implication is that the field vector always undergoes fast rolling from a region with $V>0$ towards an anti-de Sitter vacuum.

The no-go theorem we have derived here can be interpreted as defining
a necessary condition for inflation in IIA models: 
{\it In order to
inflate, a IIA compactification must contain some additional structure
beyond that considered so far which gives a term in the potential $V$
whose scaling leads to a term on the RHS of (\ref{operator}) with a
coefficient less than 9 if positive or greater than 9 if negative}.  In the next section we turn to a discussion
of specific types of structure which can realize this necessary
condition for inflation.

\section{Evading the No-Go Theorem}\label{Evade}

In the analysis which led to the preceding no-go theorem for inflation
in IIA compactifications, we allowed a specific set of ingredients in
the IIA models considered.  Following \cite{DGKT}, we included NS-NS
3-form flux, R-R fluxes, D6-branes and O6-planes, which are sufficient
to stabilize all geometric moduli.  Because D6-branes and
anti-D6-branes give terms to $V$ which scale in the same way, this no-go 
theorem rules out brane-antibrane inflation as well as inflationary
models using other moduli as the inflaton.  Corrections to
10-dimensional supergravity which arise at small compactification
volume may evade our no-go result, so one approach to finding a IIA
compactification with inflation is to include finite volume
corrections to the K\"ahler potential.  There are also other structures
we can include in a compactification besides those already mentioned
which can evade the no-go result.  In this section we consider other
possibilities which give rise to terms in the potential $V$ which have
scaling coefficients violating (\ref{operator}).  This may give some
guidance for where to look to find IIA compactifications with
inflation and/or de Sitter vacua.  Note that many of the structures
suggested go beyond the range of compactifications which are so far
well understood in string theory.

\subsection{Various Ingredients}\label{Ingredients}

One obvious possibility is to include D$p$-branes and O$p$-planes of
dimensionality other than $p=6$.  
The scaling of the resulting contributions to $V$  are as follows
\begin{eqnarray}
V_{\mbox{\tiny{Dp}}}\amp\propto\amp \volu^{\frac{p-6}{2}}\dil^{-3}\,\,\,\,\,\,\,\,\,\,\,\mbox{for D$p$-branes},\nonumber\\
V_{\mbox{\tiny{Op}}}\amp\propto\amp -\volu^{\frac{p-6}{2}}\dil^{-3}\,\,\,\,\,\,\,\mbox{for O$p$-planes}.
\end{eqnarray}
In these cases, the right hand of eq.~(\ref{operator}) is
$(12-p/2)V_{\mbox{\tiny{Dp/Op}}}$, so the no-go theorem applies to
D$p$-branes with $p\le 6$ and O$p$-planes with $p\ge 6$, but is evaded
otherwise.  Since the branes must extend in all non-compact directions
of space-time\footnote {Otherwise they would describe localized
excitations in an asymptotic vacuum not including these branes.} in
IIA we can only consider $p = 4,\,8$. Wrapping such a brane, however,
would either require a compactification with non-trivial first
homology class $H_1$ or finite $\pi_1$,
or perhaps require the use of so-called coisotropic 8-branes
\cite{coisotropic1,coisotropic2}.  These branes carry charge and so
would generate additional tadpoles which would need to be cancelled,
as well as potentially breaking supersymmetry.  Compactifications with
branes of this type have not been studied extensively in the string
theory literature, but it would be interesting to investigate this
range of possibilities further.

Another possibility is to include more general NS-NS fluxes, such as
geometric fluxes {\em a la} Scherk and Schwarz \cite{Scherk} or
non-geometric fluxes \cite{Wecht}.  Geometric fluxes parameterize a
``twisting'' away from Calabi-Yau topology, generalizing the notion of
a twisted torus \cite{Scherk, km}.  These fluxes are associated
with a metric with curvature on the compact space.  Geometric fluxes
arise under T-duality or mirror symmetry when an NS-NS 3-form $H$ has
a single index in a dualized direction \cite{KSTT}.  Further T-dualities generate
non-geometric fluxes, described in \cite{Shelton} through the sequence
$T:H_{abc} \rightarrow f^a_{bc} \rightarrow Q^{ab}_c \rightarrow
R^{abc}$, where $f$ parameterize geometric fluxes, $Q$ are locally
geometric but globally non-geometric fluxes which can in some cases be
realized in the language of ``T-folds'' \cite{dh,Hull}, and $R$
parameterize fluxes associated with compactifications which are
apparently not even locally geometric.  These general NS-NS fluxes and
the associated compactifications are still poorly understood.  From
the T-duality picture, however, it is straightforward to determine the
scalings of these 3 types of fluxes.  Each T-duality inverts the size
of a dimension of the compactification, replacing a factor of
$\voluume^{-1}$ in the scaling with $\voluume$, so we have
\begin{eqnarray}
V_{\mbox{\tiny{f}}} \amp \propto \amp \pm\volu^{-1}
\dil^{-2}\,\,\,\,\,\,\,\mbox{for geometric (f) flux},\nonumber\\
V_{\mbox{\tiny{Q}}} \amp \propto \amp \pm\volu\,\dil^{-2}\,\,\,\,\,\,\,\,\,\,\,\,\mbox{for Q flux},\nonumber\\
V_{\mbox{\tiny{R}}} \amp \propto \amp \pm\volu^{3} \dil^{-2}\,\,\,\,\,\,\,\,\,\,\mbox{for R flux}.
\end{eqnarray}
Applying the linear operator of eq.~(\ref{operator}) to each of these
terms gives a right hand side of
$7V_{\mbox{\tiny{f}}},\,5V_{\mbox{\tiny{Q}}}$, and
$3V_{\mbox{\tiny{R}}}$, respectively.  So although our no-go theorem
applies when such terms are negative, it is evaded if any of these
contributions are positive.  Among these fluxes, the best understood
are geometric fluxes, which are realized in many simple
compactifications such as twisted tori \cite{vz,Scherk, km}.
Compactification on spaces with these fluxes (and other ingredients)
is studied in the forthcoming paper by Silverstein \cite{Eva}, where
it is shown that de Sitter vacua can indeed be realized in such
backgrounds.  This is a promising place to look for string inflation
models.
Note, however, that general NS-NS fluxes cannot in general be taken to
the large volume limit.  For example, fluxes of the $Q$ type involve a
T-duality inverting the radius of a circle in a fiber when a circle in
the base is traversed.  Thus, somewhere the size of the fiber must be
sub-string scale.  This makes solutions of the naive 4-dimensional
supergravity theory associated with flux compactifications such as
those found in \cite{stw2} subject to corrections from winding modes
and also to uncontrolled string theoretic corrections if curvatures
become large.

Another possible ingredient which can be added to the IIA
compactification models are NS 5-branes; these are the magnetic duals
of the string.  Such objects are non-perturbative and carry tension in
10 dimensions that scales as $g_s^{-2}=e^{-2\phi}$.  
While they backreact more significantly than D-branes and so are not
as simple to describe at the 
supergravity level, their presence can be captured by adding a term of
the form
\begin{equation}
-\mu_5\int_{\mbox{\tiny{NS5}}}d^6\xi\sqrt{-g}\,e^{-2\phi}
\end{equation}
 to the action of eq.~(\ref{gravity1}).  The compactification and
conformal transformation yields a new term that scales as
\begin{equation}
V_{\mbox{\tiny{NS5}}} \propto \volu^{-2} \dil^{-2},
\end{equation}
yielding a right hand side of eq.~(\ref{operator}) of $8V_{\mbox{\tiny{NS5}}}$, hence evading the no-go theorem.
To satisfy tadpole cancellation, one would probably wish to add metastable pairs of separated NS 5-branes and
anti-NS 5-branes, wrapping distinct isolated curves in the same homology class.  Such configurations have been a focus of study in the dual IIB theory in recent works, starting with \cite{Aganagic}.

\subsection{An Illustration}
In this section we illustrate how the above ingredients may be useful from the point 
of view of building de Sitter vacua and inflation.  We 
focus the discussion on NS 5-branes, which appear particularly promising.  We 
will not attempt an explicit construction, since that would take us beyond the scope of this article.  
Our goal is only to show that simple, available ingredients in the IIA theory have energy
densities which scale with the volume and dilaton moduli in a way which suffices to overcome our no-go theorem, which was based purely on scalings of energy densities.
This should act as a guide to model building, but should be taken in the heuristic spirit it is offered.


Using the potential of eq.~(\ref{potential}) and adding to it a (necessarily) positive term from NS 5-branes wrapping 2-cycles, we obtain\footnote{We have set $A_{\mbox{\tiny{D6}}}=0$ as it adds little to the analysis.}
\begin{equation}
V=
\frac{A_3(\md)}{\volu^3\dil^2}+\sum_p\frac{A_p(\md)}{\volu^{p-3}\dil^4}
-\frac{A_{\mbox{\tiny{O6}}}(\md)}{\dil^3} +\frac{A_{\mbox{\tiny{NS5}}}(\md)}{\volu^2\dil^2}.
\label{toy}\end{equation}
Let us now streamline $V$ and focus on the most important features of this setup;
we set $A_2=A_6=0$, and expect the remaining coefficients to scale with fluxes and numbers of planes and branes as 
\begin{eqnarray}
&& A_3\sim h_3^2,\,\,\, A_0\sim f_0^2,\,\,\, A_4\sim f_4^2,\,\,\,
A_{\mbox{\tiny{O6}}}\sim N_{\mbox{\tiny{O6}}},
\,\,\,
A_{\mbox{\tiny{NS5}}}\sim g(\omega) N_{\mbox{\tiny{NS5}}}, 
\end{eqnarray}
where we have introduced a function $g$ of the modulus $\omega$ associated with the 2-cycle wrapped by each NS 5-brane. 

There is a tadpole constraint that the charge on the O6-plane must be balanced by the fluxes from $H_3$ and $F_0$, i.e., $h_3 f_0 \sim - N_{\mbox{\tiny{O6}}}$. We use this to eliminate $h_3$.
Now since it has no effect on the kinetic energy, let us rescale our fields as: $\volu \to \volu\sqrt{|f_4/f_0|}$ 
and $\dil \to \dil\sqrt{|f_0 f_4^{3}|}/N_{\mbox{\tiny{O6}}}$.  Then we have
\begin{equation}
V=V_{\mbox{\tiny{flux}}}\!\left[\frac{B_3(\md)}{\volu^3\dil^2}+\sum_p\frac{B_p(\md)}{\volu^{p-3}\dil^4}
-\frac{B_{\mbox{\tiny{O6}}}(\md)}{\dil^3} +\frac{B_{\mbox{\tiny{NS5}}}(\md)}{\volu^2\dil^2}\right]
\label{toy2}\end{equation}
with $V_{\mbox{\tiny{flux}}} \equiv N_{\mbox{\tiny{O6}}}^4/\sqrt{ | f_0^3 f_4^{9} | }$, and 
\begin{eqnarray}
&& B_3\sim 1,\,\,\, B_0\sim 1,\,\,\, B_4\sim 1,\,\,\,
B_{\mbox{\tiny{O6}}}\sim 1,
\,\,\,
B_{\mbox{\tiny{NS5}}}\sim c(\omega)\equiv g(\omega) N_{\mbox{\tiny{NS5}}}\sqrt{|f_0^3 f_4|}/N_{\mbox{\tiny{O6}}}^2.
\label{comega}\end{eqnarray}
So the shape of the potential is essentially controlled by the parameter $c$.

Let us make some comments on the value of $c$ which determines the contribution of the NS 5-brane.  In the large $f_4$ flux limit of \cite{DGKT}, $c$ is parametrically larger than all other contributions, including that from the 3-form flux to which we compare:
The NS 5-brane contribution has the same scaling with $\dil$, 
but scales more slowly to zero as $\rho \to \infty$.
Therefore, one would expect that one needs ${\it small}$ $N_{\mbox{\tiny{NS5}}}$ to push $V$ up  without causing a runaway
to infinite volume.  As we will see, we need $c$ to be fined tuned and $\mathcal{O}(1)$.
An analogous situation occurs in IIB string theory with
compactifications involving anti-D3 branes and non-perturbative volume stabilization.  
There the presence of strong warping allows one to construct states where
the anti-D3 energy density is exponentially suppressed, naturally providing a small coefficient to the perturbation of
the energy density \cite{KPV}.  This plays an important role in de Sitter constructions in that context \cite{KKLT},
and one might expect an analogous mechanism (involving large warping or a very small cycle) could similarly dynamically explain a small $g(\omega)$ to compensate large $f_4$ in the IIA context.
We could also imagine a compactification where $N_{\mbox{\tiny{O6}}}$ is very large to achieve the same result.

In any case, by 
treating $c$ as a continuous parameter and ignoring the dynamics of all other moduli, we can obtain a meta-stable de Sitter vacuum.  We set ($\mpl=1$) $B_3=1/4,\,B_0=1/4,\,B_4=3/8,\,B_{\mbox{\tiny{O6}}}=2$, and stabilize $\volu$ by satisfying $\partial V/\partial\volu = 0$.
In Figure 1 we plot $V=V(\hat{\dil})$ of eq.~(\ref{toy2}) for different choices of $B_{\mbox{\tiny{NS5}}}=c$.  We find that there are critical values of $c$: for a Minkowski vacuum $c_M\approx 2.205$ and for a point of inflection $c_I \approx 2.280$.
\begin{figure}[t]
\begin{center}
  \epsfxsize = 12cm
  \centerline{\epsfbox{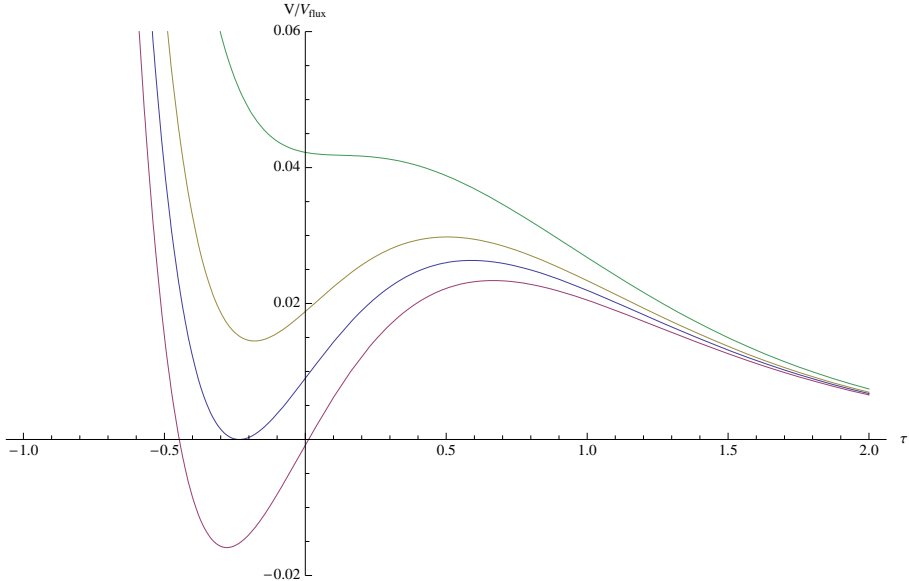}}
  \caption{The potential $V(\hat{\tau})/V_{\mbox{\tiny{flux}}}$  with ($\mpl=1$) $B_3=1/4,\,B_0=1/4,\,B_4=3/8,\,B_{\mbox{\tiny{O6}}}=2$, and $\volu$ satisfying $\partial V/\partial \volu=0$.  From bottom to top, the curves correspond to the following choices of $c$\,: 2.183 (anti-de Sitter), 2.205 (Minkowski), 2.227 (de Sitter), and 2.280 (inflection), respectively.}
\end{center}
\end{figure}
For $c<c_M$ an anti-de Sitter vacuum exists, for $c_M<c<c_I$ a de Sitter vacuum exists, and for $c>c_I$ no vacuum exists.  For $c$ close 
to $c_M^{+}$ we expect the de Sitter vacuum lifetime to be long, as in the KKLT meta-stable vacuum of type IIB \cite{KKLT}.

If one could realize such a construction explicitly in a controlled
regime of the IIA theory, then we anticipate that the
opportunities to realize inflation will be greatly enhanced. For
example, the local maximum in Figure 1 may be useful since $\epsilon\to
0$ there. It will often suffer, however, from the so-called $\eta$
problem; the second slow-roll parameter (which measures the second
derivative of the potential) will typically be large and negative.  It
is possible that, for example, by moving in another transverse
direction at the hill-top one could build some form of
hybrid-inflation.  We do note that by choosing $c=c_I$ we would
immediately solve the $\eta$ problem and have
inflection-point-inflation as advocated in \cite{Itzhaki}. In this situation,
however,
inflation would finish with two difficulties; runaway moduli and
little to no reheating.  A different approach is to simply fix the
volume and the dilaton at a de Sitter (or Minkowski) minimum with high mass
and strictly use other lighter moduli to
drive inflation in transverse directions.
Scenarios such as N-flation may be possible here \cite{axiontwo}.

\section{Type IIB Compactifications}
\label{sec:IIB}

In this section we discuss the relationship between the results we
have derived here for IIA string theory and previous work on inflation
in type IIB string theory.  Although the type IIA and IIB string
theories are related through T-duality, this duality acts in a
complicated way on many of the ingredients used in constructing flux
compactifications.  The basic IIB flux compactifications of \cite{GKP}
(see also \cite{Becker,GVW,DRS})
involve $H$-flux, $F_3$-flux, D3-branes and O3-planes.
T-duality/mirror symmetry on such compactifications transforms the
NS-NS $H$-flux into a complicated combination of $H$-flux, geometric
flux and non-geometric flux which generically violates the
restrictions needed in the no-go theorem we have proven here.
Conversely, the backgrounds which we have proven here cannot include
inflation are T-dual/mirror to complicated IIB backgrounds with
geometric and non-geometric fluxes.  Furthermore, the volume modulus
used in our analysis is dual to a complex structure modulus in the IIB
theory which is difficult to disentangle from the other moduli, so
that proving the analogous no-go theorem in IIB, on the exotic class
of backgrounds where it is relevant, would be quite difficult without
recourse to duality.

Despite these complications, we can easily explain why standard IIB
flux vacua behave so differently with respect to potential
constructions of de Sitter space and inflation as seen through the
methods of this paper.  For IIB flux vacua,
the basic ingredients of $H$-flux, $F_3$-flux, D3-branes and O3-planes
give contributions to the 4-dimensional potential which scale as
\begin{equation}
\frac{e^{2 \phi}}{\voluume^6},\,\,\,
\frac{e^{4 \phi}}{\voluume^6},\,\,\,
\frac{e^{3 \phi}}{\voluume^6},\,\,\,
-\frac{e^{3 \phi}}{\voluume^6}
\end{equation}
respectively.\footnote{This $\rho$ scales as Vol$^{1/3}$ defined in the string frame, as in Section II, and should not be confused with the $\rho$ modulus of e.g. \cite{GKP}, which scales as ${\rm Vol}^{2/3}$ defined in the Einstein frame.}
Thus, the scaling equation  (\ref{operator}) is not the
appropriate equation for gaining useful information about the cosmological
structure.  Instead, we simply have
\begin{equation}
-\voluume \frac{\partial V (\phi, \voluume)}{ \partial\voluume}  = 6V (\phi,
 \voluume) \,.
\label{eq:}
\end{equation}
This shows immediately that any classical vacuum must have $V = 0$, as is
well known from the tree-level no-scale structure of this class of models \cite{GKP}.
When the dilaton and complex structure moduli are
chosen to fix $V = 0$ then there is a classical flat direction. When
such moduli cannot be chosen then the (positive) potential causes a
runaway to large volume. 
In such a simple setting, one can show that $\epsilon\ge 3$ whenever $V>0$, but this {\em should not}
be viewed as a serious obstacle to realizing inflation or de Sitter vacua.
This is because the classical
flat directions along which $V = \partial V/\partial\voluume= 0$ can be lifted by including quantum contributions (or other fluxes etc) to the potential, and then the naive bound on $\epsilon$ is irrelevant.
Typically, non-perturbative corrections to the superpotential are included to
stabilize the K\"ahler moduli.  
This should be contrasted with the classical stabilization in IIA.\footnote{
Note that in the absence of fluxes, branes and orientifolds, mirror
symmetry relates supersymmetric IIA and IIB compactifications.  In
this case $V = 0$ exactly.}

In this general IIB setting, starting with the no-scale vacuum and including
various corrections to achieve de Sitter, many inflationary models have been proposed.  The basic 
strategy, starting with \cite{branethree},
has usually been to stabilize the dilaton and volume moduli at a high scale, and inflate at
a lower energy scale.  Then, the dilaton and volume contributions to $\epsilon$
(which were the focus of our no-go theorem in the IIA context) are simply absent.
The most explicit models to date appear in \cite{branefive}, though one should 
consult the reviews \cite{reviews1,reviews2,reviews3,reviews4,reviews5} for a much more extensive list of approaches
and references.

\section{Discussion}\label{Discussion}

In this paper we have demonstrated that a large class of flux
compactifications of type IIA string theory cannot give rise to
inflation in the regime of moduli space where we have parametric
control of the potential.  This result applies to large-volume, weak
coupling compactifications on arbitrary Calabi-Yau spaces with NS-NS
3-form flux, general R-R fluxes, D6-branes and O6-planes.  These
ingredients are arguably the most well understood in IIA
compactifications.  The no-go theorem of Section \ref{Nogo} applies in
particular to the $T^6/\mathbb{Z}_3^2$ orientifold model of
Ref.~\cite{DGKT} and the $T^6/\mathbb{Z}_4$ orientifold model of
Ref.~\cite{IW}, and explains the numerical results of
Ref.~\cite{Hertzberg} which suggested that inflation is impossible in
these models.  The no-go theorem we have derived here, however,
applies to all other Calabi-Yau compactifications of this general type
as well.  So the simplest part of the IIA flux compactification
landscape does not inflate.  This implies the following constraint: the portions of the
landscape that are possibly relevant to phenomenology will necessarily
involve interplay of more diverse ingredients, as has also been found
in the IIB theory.

We emphasize that while our derivation has only involved
two moduli (the volume $\volu$ and the dilaton $\dil$)
we are not assuming that either of those moduli necessarily play the role of the
inflaton.  Instead, inflation by {\em any} modulus (or brane/anti-brane) is always spoiled due to the fast-roll of $\volu$ and/or $\dil$. 
This follows because a necessary condition for slow-roll inflation is that the potential be flat in {\em every} direction in field space, 
as quantified by $\epsilon$.
In fact because the first slow-roll parameter is so large $\epsilon\geq \frac{27}{13}$, there can 
never be many e-foldings, even ruling out so-called {\em fast-roll inflation} \cite{Fast}.
We point out that this result is slightly non-trivial because it requires analyzing {\em both}
$\volu$ and $\dil$, as in eq.~(\ref{operator}), and cannot be proven by focussing on only one of them.

A simple corollary of our result is that no parametrically
controlled de Sitter vacua exist in
such models.  We emphasize, though, that proving the non-existence of
inflation is a much stronger statement than proving the non-existence
of de Sitter vacua.  In particular we can imagine a priori a scenario
where, although the vacuum is anti-de Sitter (or Minkowski) or there is no vacuum at all, inflation
is still realized somewhere in a region where $V>0$.
We find, however,
that this does not occur.  It is intriguing that the proof of the
non-existence of inflation is so closely related to that of the
non-existence of de Sitter.  This {\em suggests} that there may be
a close connection between building de Sitter vacua and realizing
inflation.

Our result can be interpreted as giving a necessary condition for
inflation in IIA models.  To have inflation, 
some additional structure
must be added which gives rise to a potential term in 4 dimensions
with scaling such that $-\rho \partial V/\partial \rho -3 \tau
\partial V/\partial \tau = \alpha V$ with a coefficient $\alpha < 9$
for a positive contribution and $\alpha > 9$ for a negative
contribution.  
We described various ingredients which give rise to such terms;
   compactifications including these ingredients may be promising places
   to look for inflationary models.
Some of these ingredients take us outside
the range of string compactifications which are understood from a
perturbative/supergravity point of view.  Among the possibilities
which evade the assumptions made in deriving the no-go theorem are
other NS-NS fluxes, such as geometric and non-geometric fluxes.  It is
currently difficult to construct models with such generic fluxes in a
regime that is under control, but progress in this direction has been
made in \cite{Eva}, where IIA de Sitter vacua are found using a
specific set of geometrical fluxes and other ingredients.  Another
promising direction which we have indicated here (also incorporated in
\cite{Eva}) is to include NS 5-branes and anti-NS 5-branes on 2-cycles in
the Calabi-Yau.  More work is needed to find explicit models where
these branes are stabilized in a regime allowing de Sitter vacua and
inflation, but this does not seem to be impossible or ruled out by any
obvious considerations.  In addition to the mechanisms we have discussed,
there are probably other structures (e.g., D-terms \cite{Wrase2,Wrase3})
which violate the conditions of the no-go theorem.

Including any of these ingredients does not guarantee that
inflation will be realized.  It may be the case that a slightly more general
no-go theorem for inflation exists with certain combinations of additional ingredients. 
This may follow from studying other moduli, since any field can ruin inflation by fast-rolling. 
It may also be that with
more work an elegant realization of inflation and  de Sitter
vacua can be found in type IIA string theory.  This deserves further investigation.

\acknowledgments 
We would like to thank 
Thomas Grimm, Hong Liu, John McGreevy, Mike Mulligan, Eva Silverstein, and David Vegh
for comments on a preliminary version of this paper and for helpful discussions.
Thanks also to Marcus Berg, Ulf Danielsson, Vijay Kumar, and Frank Wilczek for helpful
discussions. We would also like to thank Eva Silverstein for sharing with us an
advance copy of her work \cite{Eva}.  M.~P.~H.  acknowledges the STINT foundation and the
kind hospitality of the Department of Theoretical Physics at Uppsala University where part of this work was carried out.  
%
%
M.\ P.\ H. and W.\ T.\ thank the Department of
Energy (D.O.E.) for support under cooperative research agreement
DE-FC02-94ER40818.  S.\ K.\ acknowledges the kind hospitality of the
MIT Center for Theoretical Physics during the initiation of this work,
and the NSF and DOE for support under contracts PHY-0244728 and
DOE-AC02-76SF00515. S.\ K.,
M.\ P.\ H., and M.\ T.\ acknowledge support from 
NASA grant NNG06GC55G, NSF grants 
AST-0134999, 0607597 \& 6915954, and 
the David and Lucile Packard Foundation.

\appendix	

\section{4-Dimensional ${\cal N}=1$ Supergravity}\label{App}

It was shown in \cite{Grimm} that with the ingredients used in Section
\ref{Nogo}, the dimensional reduction of massive IIA supergravity can
be described in the language of a 4-dimensional ${\cal N} = 1$ supergravity
theory in terms of a K\"ahler potential and superpotential.  

\subsection{Kinetic Energy from K\"ahler Potential}\label{AppK}

The authors of Ref.~\cite{Grimm} showed that in the large volume limit the
K\"ahler potential is given by $K=K^K+K^Q$ where
\begin{eqnarray}
K^K 
& = & -\mpl^2\ln\left(\frac{4}{3}\kappa_{abc}v^av^bv^c\right),\,\,\\
K^Q 
& = & -2\mpl^2\ln\Big{(}2\,\mbox{Im}(C Z_\lambda)\mbox{Re}(Cg_\lambda)
-2\,\mbox{Re}(CZ_k) \mbox{Im}(Cg_k)\Big{)}\label{KQ}
\end{eqnarray}
Here $v^a$ are the K\"ahler moduli, $\kappa_{abc}$ are the triple-intersection form constants, 
the set of $Z$ and $g$ are the co-ordinates in some basis of a 
holomorphic 3-form that describes the complex structure moduli,
and $C$ is the ``compensator" which incorporates the dilaton. If the set of complex moduli is denoted $\psi^i$,
then the kinetic energy is given by
\begin{equation}
T = - K_{i\bar{j}}\partial_\mu\psi^i\partial^\mu\psi^{\bar{j}}
\end{equation}
with corresponding first slow-roll parameter
\begin{equation}
\epsilon = \mpl^2 \frac{K^{i\bar{j}}V_i V_{\bar{j}}}{V^2}.
\label{srfull}\end{equation}

Let us focus on the K\"ahler contribution. We write the K\"ahler moduli as $\psi^a=a^a+i\,v^a$, so the kinetic energy is given by
\begin{equation}
T^K = -\frac{1}{4}\frac{\partial^2 K^K}{\partial v^a\partial v^{b}}\left(\partial_\mu v^a\partial^\mu v^{b}
+\partial_\mu a^a\partial^\mu a^{b}\right).
\end{equation}
Now we change coordinates from $v^a$ to $\{\rho,\gamma^a\}$ as follows:
\begin{equation}
v^a=\rho\,\gamma^a,\,\,\,\mbox{with}\,\,\,\kappa_{abc}\gamma^a\gamma^b\gamma^c=6,
\end{equation}
so Vol\,=\,$\volu^3$.
Then using $\partial_\mu(\kappa_{abc}\gamma^a\gamma^b\gamma^c)=0$, we obtain:
\begin{eqnarray}
T^K \amp=-\amp\mpl^2\!\Bigg[\frac{3(\partial_\mu\volu)^2}{4\,\volu^2}-\frac{1}{4}\kappa_{abc}\gamma^c\partial_\mu\gamma^a\partial^\mu\gamma^b
+\frac{\kappa_{acd}\gamma^c\gamma^d\kappa_{bef}\gamma^e\gamma^f
- 4\kappa_{abc}\gamma^c}{16\,\volu^2}
\partial_\mu a^a\partial^\mu a^b
\Bigg]\,\,\,\,\,\,\,\,\,
\end{eqnarray}
By switching from $\volu$ to $\hat{\volu}$, we see that the first term is precisely the kinetic energy for $\hat{\volu}$. The remaining kinetic energy terms for $\gamma^a$ and $a^a$ are block diagonal (there are no cross terms involving $\partial_\mu\volu\,\partial^\mu\gamma^a$ etc), and this has an important consequence:
We know that in the physical region the total kinetic energy must be positive, so  {\em each} of the above 3 terms must be positive. Hence, $T^K=-(\partial_\mu\hat{\volu})^2/2+\mbox{positive}$.

For the complex structure/dilaton sector the procedure is similar, although more subtle.
In (\ref{KQ}) the expression for $K^Q$ is not a completely explicit function of the moduli;
although $\mbox{Re}(CZ_k)$ and $\mbox{Re}(Cg_k)$ are explicitly half of the complex structure moduli, $\mbox{Im}(CZ_k)$ and $\mbox{Im}(Cg_k)$ are only {\em functions} of the remaining complex structure moduli. 
Nevertheless, the kinetic term is again block diagonal.
To see this, note that the compensator $C$, and hence all moduli in this sector, are proportional to 
$\dil$. Furthermore, the $Z$ and $g$ are constrained to the surface: 
$K^Q=-2\mpl^2\ln(\dil^2)$. This is analogous to the K\"ahler sector. Without going through the details here, we find $T^Q=-(\partial\hat{\dil})^2/2+\mbox{positive}.$
In fact we know this {\em must} be true from the 10-dimensional point of view; the dilaton modulus is inherited directly from 10 dimensions, and so cannot possibly give rise to mixed kinetic terms with the complex structure moduli in 4 dimensions.

\subsection{Potential Energy from Superpotential}\label{AppW}

From \cite{Grimm} the superpotential in the IIA theory is given by $W=W^K+W^Q$ where
\begin{eqnarray}
W^K & = & f_6+f_{4a}t^a+\frac{1}{2}f_{2a}\kappa_{abc}t^bt^c-\frac{f_0}{6}\kappa_{abc}t^at^bt^c,
\,\,\,\,\,\,\,\,\,\,\, \\
W^Q & = & \left(\bar{h}_\lambda \xi^\lambda-h_k\xi^k\right) 
+2i\left(\bar{h}_\lambda\mbox{Re}(Cg_\lambda)
-h_k\mbox{Re}(CZ_k)\right).
\end{eqnarray}
We will not explain all the details of this here; the interested reader is pointed to Refs.~\cite{Grimm,DGKT}. For our present purposes it suffices to
note that Im$(t^a)=v^a\propto\volu$ and Im$(W^Q)\propto\dil$.
Hence, the superpotential is cubic in 
$\volu$ and linear in $\dil$.  

From the supergravity formula for the Einstein frame potential 
\begin{equation}
V=e^{K/\mpl^2}\left(D_i W K^{i\bar{j}} \overline{D_j W}-3\frac{|W|^2}{\mpl^2} \right),
\label{SUGRA}\end{equation}
we easily infer the dependence on $\volu$ and $\dil$.  Firstly, since the constraints on $\gamma^a$
and complex structure imply that $K=-\mpl^2\ln(8\volu^3\dil^4)$,
the pre-factor scales as
\begin{equation}
e^{K/\mpl^2}\propto \volu^{-3}\dil^{-4}.
\end{equation}
Also, the scaling contributions from the parenthesis in eq.~(\ref{SUGRA}), which is roughly $|W|^2$, can 
be easily determined. By analyticity, only terms of the form $\volu^p\dil^q$ can appear where $p+q$ is even.
This leaves only the following 7 possible scalings: $\dil^2, \volu^6, \volu^4, \volu^2, 1, \volu^3\dil, \volu\dil$.
By multiplying by the pre-factor, we see that the first 6 terms are precisely those that arise from $H_3, F_0, F_2, F_4, F_6$, and D6/O6, respectively.  The 7th term ($\volu\dil$) is new, but cancels between the two terms inside the parenthesis of eq.~(\ref{SUGRA}).  Hence we obtain the form of the potential given in eq.~(\ref{potential}).

\end{document}